\documentclass[aps,pra,twocolumn,10pt,groupedaddress,nofootinbib,superscriptaddress,notitlepage,showpacs,floatfix]{revtex4-1}
\usepackage{graphicx,graphics}
\usepackage{epstopdf}
\usepackage{subfigure}
\usepackage{dcolumn}   
\usepackage{array}
\usepackage{amsthm}
\usepackage{amsmath}
\usepackage{amssymb}
\usepackage{amsfonts}
\usepackage{mathrsfs} 
\usepackage{bm}
\usepackage{braket} 
\usepackage{enumitem}
\usepackage{url}
\usepackage[pdfstartview=FitH,pdfencoding=auto]{hyperref}
\hypersetup{
    colorlinks=true,        		
    linkcolor=blue,              	
    citecolor=blue,        			
    filecolor=blue,          		
    urlcolor=blue,              	
    runcolor=cyan
			}
\usepackage[pdftex]{color}
\usepackage{multirow}
\usepackage{CJK}

\newcommand{\ud}{\mathrm{d}}

\begin{document}
\begin{CJK*}{GB}{}
\title{Feedback Ansatz for Adaptive-Feedback Quantum Metrology Training with Machine Learning}
\author{Yi Peng}
	\affiliation{Institute of Physics, Chinese Academy of Sciences, Beijing 100190, China}
	\affiliation{School of Physical Sciences, University of Chinese Academy of Sciences, Beijing 100190, China}
\author{Heng Fan}
	\email{hfan@iphy.ac.cn}
	\affiliation{Institute of Physics, Chinese Academy of Sciences, Beijing 100190, China}
\affiliation{School of Physical Sciences, University of Chinese Academy of Sciences, Beijing 100190, China}
	\affiliation{CAS Center for Excellence in Topological Quantum Computation, University of Chinese Academy of Sciences, Beijing 100190, China}
	\affiliation{Songshan Lake Materials Laboratory, Dongguan 523808, Guangdong, China}
\date{\today}
\begin{abstract}
	It is challenging to construct metrology schemes which harness quantum features such as entanglement and coherence to surpass 
	the standard quantum limit. We propose an ansatz for devising adaptive-feedback quantum metrology (AFQM) strategy which greatly reduces the
	searching space. Combined with the Markovian feedback assumption, the computational
	complexity for designing AFQM would be reduced from $N^7$ to $N^4$, for $N$ probing systems.
	The feedback scheme devising via machine learning such as particle-swarm optimization and differential evolution {would thus}
	require much less time and produce equally good imprecision scaling. We have thus devised an AFQM for $207$-partite system.
	The imprecision scaling would persist for $N>207$ in an admirable range  when the parameter setting for $207$-partite system is employed without
	further training.
	{Our ansatz indicates an built-in resilience of the feedback strategy against qubit loss. The feedback strategies designed for 
	the noiseless scenarios have been tested against the qubit loss noise and the phase 
	fluctuation noise. Our numerical result confirms great resilience of the feedback strategies against the two kinds of noise.}
\end{abstract}
\maketitle
\end{CJK*}
\section{Introduction}
	Given $N$ entangled probing systems, quantum metrology promises  parameter estimation with imprecision
	below the lowest limit $1/\sqrt{N}$ allowed by classical theory. This is known as the
	{\it standard quantum limit (SQL)}. The lowest imprecision permitted by quantum mechanics is $1/N$, i.e. the
	so-called {\it Heisenberg limit (HL)}~\cite{braunstein1994statistical,giovannetti2004quantum-enhanced,
	giovannetti2006quantum,giovannetti2011advances,Degen2017,Pezze2018}.  Such kind of quantum superiority over the classical
	schemes attracts much attention in both academic and industry communities. Because it has wide range of  applications
	including spectroscopy~\cite{Bollinger1996}, accurate clock
	construction~\cite{Kessler2014,Derevianko2011,Zhang2016}, gravitational wave detection~\cite{LIGO2013,Schnabel2010}, fundamental
	biology research and medicine development~\cite{Taylor2016}, and others~\cite{giovannetti2004quantum-enhanced,Demkowicz-Dobrzanski2015,Degen2017,Pezze2018}.
	
	There are at least three prominent challenges in practical quantum metrology realization.
	a) Both SQL and HL are asymptotic and require great amount of data to approach.
	It is a serious limitation in many circumstances.  For instance, the gravitation detection window is very
	narrow~\cite{LIGO2013,Schnabel2010} while many biological samples are too fragile to endure much photon
	bombardment~\cite{Taylor2013,Taylor2016}. Thus, we need to finish the interference in limited time and with limited number of probing
	systems. b) Environment noise which is inevitable in practical platforms can completely demolish such quantum
	advantage~\cite{Escher2011general,Demkowicz-Dobrzanski2012,Haase2016}. c) Many metrology schemes proposed require
	input states or final measurements which are difficult to realize. For example, the
	Greenberger-Horne-Zeilinger (GHZ) state has ability to asymptotically achieve
	HL~\cite{Leibfried2004,giovannetti2004quantum-enhanced,giovannetti2006quantum,giovannetti2011advances,Monz2011,Liu2015}.
	Synthesising GHZ state
	is well recognized as highly complicated and inefficient~\cite{Monz2011,Wang2018,Zhang2018,Wei2006,Barends2014,Song2019,Omran2019,Wei2019}.
	In typical phase estimation tasks, canonical positive-operator-valued measure (POVM) based on the so-called \emph{phase state} $\ket{\phi}\bra{\phi}$ and the sine input state
	(\ref{sine_state_def}) have been frequently utilized to demonstrate asymptotic HL~\cite{Sanders1995,Wiseman1995,DAriano1998,VanDam2007,Hassani2017}. 
	{ One can see that the definitions of the phase state and sine state is mathematically elegant 
	\begin{equation}
		\ket{\phi} = \sum_{\mu=-j}^j\frac{e^{\mu\phi}\ket{j\mu}_y}{\sqrt{2j+1}},
		\ket{\psi_\mathrm{\sin}}
		= \sum_{\mu=-j}^{j}\frac{\sin\left\lbrack{\frac{(\mu+j+1)\pi}{2(j+1)}}\right\rbrack\ket{j\mu}_y}{\sqrt{j+1}}.
		\label{sine_state_def}
	\end{equation}
	The physical background is not clear. Here $j=N/2$ and $\ket{j\mu}_{x,y,z}$ is the eigenstate of $\hat{J}_{x,y,z}$ respectively, belonging to eigenvalue 
	$\mu$.}
	To our knowledge, there is no clear way to realize either of them for
	$N\ge3$.

\begin{figure}[ht]
		\includegraphics[width=0.45\textwidth]{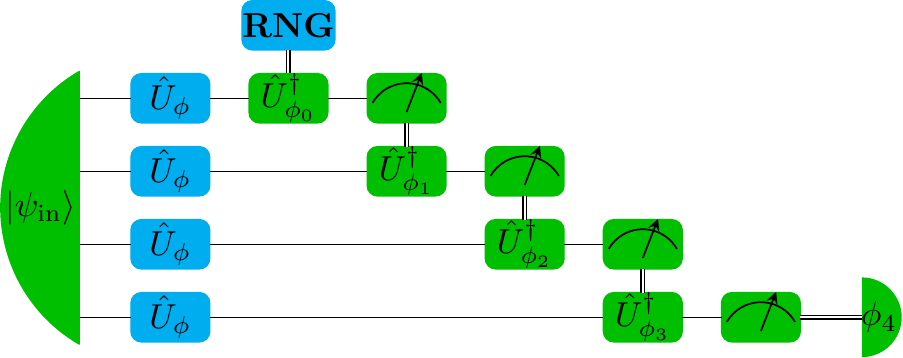}
		\caption{(Color online) Quantum circuit of AFQM employing local projective measurement for $N=4$. $\ket{\psi_\mathrm{in}}$ is the total input state.
		The interference process
		$\hat{U}_\phi$ is controlled by $\phi$. $\phi_0$ is the initial random guess generated by a random number generator (RNG).
		$\phi_1$ is feedback information gathered from the first  measurement, $\phi_2$ from the
		first and second measurements while $\phi_3$ from the first three measurements.
		$\phi_4$ is the final single-shot estimation of $\phi$ determined by all the measurements. Case of arbitrary $N$ is
		similar.} \label{feedback_qmetr_circ}
	\end{figure}

	The adaptive-feedback quantum metrology (AFQM) is believed to be a promising candidate capable of giving good parameter
	estimation
	with limited number of measurements and thus resolve issue a). As an example, the so-called Berry-Wiseman-Breslin scheme
	(BWB) can provide single-shot estimation achieving imprecision below SQL. Besides, BWB employs local projective
	measurements which partially resolves issue c)~\cite{berry2000optimal,berry2001optimal}. BWB is a
	well-educated heuristic strategy. Devising AFQM is highly challenging. Considering the AFQM employing local projective
	measurements as described in Fig.~\ref{feedback_qmetr_circ}, the total measurement outcome combinations as well as the feedbacks would
	amount to $2^N$ if $N$ qubits are employed. It indicates plenty flexibility of this type of AFQM scheme as well as a great challenge of
	optimizing it.
	Hentschel and Sanders firstly proposed the Markovian feedback assumption which reduces the dimension of the feedback parameter space to
	$N$. They showed that promising AFQM can autonomously devised via machine learning such as particle-swarm optimization (PSO) and 
	differential evolution (DE). We call such a scheme devising procedure as the Hentschel-Sanders approach
	(HS)~\cite{Hentschel2010,Hentschel2011}. If the noise is absent, the achievable imprecision scaling breaches SQL
	and shows superiority over BWB (cf. Table~\ref{sum_previous_results}).	 	
	\begin{table}[hb!]
		\caption{Summary of previous results. $\alpha$ is the inverse-scaling power of the imprecision $\delta\phi$ with respect to $N$.
		$N_\mathrm{max}$ is the maximum prob number of which AFQM can be obtained via HS approach.}
		\begin{ruledtabular}
			\begin{tabular}{lllllllll}
				&\multicolumn{2}{c}{BWB}
				&\multicolumn{2}{c}{PSO}
				&\multicolumn{4}{c}{DE}
				\\ \cline{2-3}\cline{4-5} \cline{6-9}
				Ref.
				&\cite{Hentschel2010}
				&\cite{Hentschel2011}
				&\cite{Hentschel2010}
				&\cite{Hentschel2011}
				&\cite{Lovett2013}	
				&\cite{Palittapongarnpim2016}
				&\cite{Palittapongarnpim2017b} 	
				&\cite{Palittapongarnpim2019a}\\
				$\alpha$	
				&$0.704$
				&$0.708$
				&$0.736$	
				&$0.747$	
				&$0.74$	
				&$0.71$						
				&$0.7198$	
				& $0.729$
				\\
				$N_\mathrm{max}$
				&&
				&$\le14$
				&$\le50$
				&$\le98$
				&$\le100$
				&$\le100$
				&$\le100$
			\end{tabular}
		\end{ruledtabular}
		\label{sum_previous_results}
	\end{table}
	Given permutation symmetric input state, schemes thus devised can provide single-shot estimation and have remarkable resilience against 
	noise. Though only the sine and product input states have been considered, HS can be applied to other states.
	Hence HS approach can solve a), b) and c) simultaneously.
	Generating such an AFQM  would consume $\mathcal{O}(N^7)$ time for computation and require $\mathcal{O}(N)$ memory space for storing the
	AFQM policy.
	AFQM for up to $N=100$ has been devised~\cite{Hentschel2010,Hentschel2011,Hentschel2011b,Lovett2013,Palittpongarnpim2016,Palittapongarnpim2017,Palittapongarnpim2017b,Palittapongarnpim2019,Hayes2014}.
	{Recently, an experiment has been conducted of implementing the HS approach on devising adaptive feedback scheme for up to $40$ single photons in product state.
	Its robustness against noise has also been shown~\cite{Lumino2018}.}

 Here we introduce an ansatz for devising AFQM which aims to tune the feedback adjustment to the sensitivity of the corresponding probing 
 systems measured before the very feedback. 
 It can reduce the feedback  space dimension from $N$ to a chosen constant, if we further adopt the Markovian 
 feedback assumption. The memory space for storing policy would also be constant. As a result, we can achieve persistent imprecision scaling for
 up to $N=207$ without increasing the training time for big $N$.  Further, we can generate an $N$-partite scheme without knowing schemes for
 fewer qubits which is required in the HS approach~\cite{Hentschel2010,Hentschel2011,Hentschel2011b,Lovett2013,Palittpongarnpim2016,Palittapongarnpim2017,Palittapongarnpim2017b,Palittapongarnpim2019,Hayes2014}.
 The computation time thus scales as $\mathcal{O}(N^4)$. We test our ansatz for devising
 AFQM via PSO as well as DE. Both  the previously studied sine state (\ref{sine_state_def}) and the spin-squeezed
 state (SSS) are considered. SSS is widely believed to have great resilience against 
 noise~\cite{sorensen2001many,Dunningham2002,ma2011quantum,Duan2011,Zhang2013,Pezze2013} and its synthesis has been realized in many
labs~\cite{Hald1999,Fernholz2008,Takano2009,Gross2010,Leroux2010,Hamley2012,Sewell2012,Muessel2014,Hosten2016,Zou2018}.
 The performance of AFQM thus devised is as good as the performance of the AFQM  devised via HS. One of the most intriguing part
 is that when applying the AFQM feedback policy obtained for $207$-partite system to bigger systems $N>207$ without further training, the imprecision 
 scaling persists in an admirable range of $N>207$.
 {Our ansatz describes a decreasing adjustment of the phase feedback with respect to each step. As a result, the phase 
 compensations near the end are also good estimations of the real parameter. It brings a built-in resilience against qubit loss noise from the feedback
 sequence. We tested the feedback policy obtained for noiseless metrology against qubit loss as well as phase fluctuations noise. Its resilience against
 the two types of noise has been confirmed by our numerical data.}

%
\section{Feedback ansatz for AFQM.}
Given $N$ spin-$\frac{1}{2}$ probes, the interference process is characterized by $\phi$
	\begin{equation}
		\hat{U}_\phi = e^{-i\phi\hat{J}_y},\quad\textrm{with}\quad
		\hat{J}_y = \sum_{n=1}^N\hat{s}_y^{(n)}.
	\end{equation}
	$\hat{J}_{x,y,z}$ denote total angular momentum along the $x$, $y$ and $z$ direction respectively while $\hat{s}_{x,y,z}^{(n)}$
	are spin operators of the $n$th probe.
	After the $n$th probe has passed through the parameter channel $\hat{U}_\phi$, we apply feedback $\hat{U}_{\phi_{n-1}}^\dag$ to 
	compensate $\hat{U}_\phi$ as closely as possible. The initial compensation $\phi_0$ is a random guess between $-\pi$ and $\pi$. Note that we
	assume $\phi\in[-\pi,\pi)$. Then we measure $\hat{s}_z^{(1)}$, the result of which would be used to adjust the next compensation $\phi_1$. The compensation-measurement-adjusting procedure carries on
	until we obtain the final estimation $\phi_N$. The $n$th compensation $\phi_n$ can be regarded as an update of $\phi_{n-1}$ with an
	adjustment determined by the $n$ previous  measurement outcomes $s_1,\ldots,s_n$ of $\hat{s}_z^{(1)},\ldots,\hat{s}_z^{(n)}$
	\begin{equation}
		\phi_n = \phi_{n-1} - \Delta_n(s_1,\ldots,s_n).
	\end{equation}
	{Note that the measurement result $s_n$ is single-shot result in every step instead of being an 
	ensemble average.}
	We want $\Delta_n(s_1,\ldots,s_n)$ to bring $\phi_n$ closer to $\phi$ in each step and $|\phi_n-\phi|$ decreases with respect to $n$.
	To achieve that, one needs to tune the feedback adjustment $\Delta_n(s_1,\ldots,s_n)$ to the sensitivity of the $n$ probes measured.
	We cannot allow $\Delta_n(s_n,\ldots,s_n)$ being too big compared with $|\phi_{n-1}-\phi|$. Because big adjustment means big fluctuation of 
	$\phi_{n}$ which would likely lead to bigger $|\phi_n-\phi|$ than $|\phi_{n-1}-\phi|$ and poor estimation in the end.
	Neither can $\Delta_n(s_n,\ldots,s_n)$ be too small. This is due to the fact the sensitivity of the system is limited by SQL and HL. 
	An adjustment way lower than  HL would not likely be felt by the system and thus would not be of much effect. Consider the case when 
	$\Delta_n(s_n,\ldots,s_n)=s_n/2^{n-1}$. $\Delta_n(s_n,\ldots,s_n)$ would be relatively too big given $n$ is small while too small when 
	$n$ is big. If the measurement result $s_n$ makes $\Delta_n(s_n,\ldots,s_n)$ move $\phi_n$ away from $\phi$, the best of all the later
	adjustments $\Delta_{n'}(s_n,\ldots,s_n)$ with $n'>n$ can achieve is to neutralize the detrimental effect of
	$\Delta_n(s_n,\ldots,s_n)$. In such a circumstance, the final estimation $\phi_N$ would be worse than $\phi_n$. It seems setting
	$\Delta_n(s_n,\ldots,s_n)$ around the order of $|\phi_n-\phi|$ would be reasonable. One can regard $\phi_1$,\ldots, and 
	$\phi_{N}$ as a serial of estimations of $\phi$. We would expect $|\phi_n-\phi|$ to be of the
	order of $1/n^\alpha$ with $\alpha$ being some positive constant between $1/2$ and $1$. $\alpha=1/2$ corresponds to SQL while
	$\alpha=1$ to HL. Another fact should be noted is that $|\phi_{n}-\phi|$ and $|\phi_{n+1}-\phi|$ are about the same order.
	Thus $\Delta_n(s_n,\ldots,s_n)$ should be smaller than $|\phi_n-\phi|$. Based on these intuitions, our feedback ansatz is
	\begin{equation}
		\Delta_n(s_n,\ldots,s_n) \propto 1/(n+1)^{\wp_n(s_1,\ldots,s_n)}.
		\label{feedbackAnsatz}
	\end{equation}
	where $\wp_n(s_1,\ldots,s_n)$ can be out of the range $[1/2,1]$ bounded by by SQL and HL.
	We used $1/(n+1)^{\wp_n(s_1,\ldots,s_n)}$ instead of $1/n^{\wp_n(s_1,\ldots,s_n)}$ to ensure that the variation of
	$\wp_1(s_1,\ldots,s_n)$ matters. So far the ansatz (\ref{feedbackAnsatz}) can only reduce the volumn of the AFQM parameter
	space. Combined with the Markovian assumption, it can reduce the parameter space drastically as shown in the following.

	{
	\section{Indications of the ansatz.}
	We elaborate two indications we can draw from ansatz (\ref{feedbackAnsatz}). The first one concerns the required property of the input state
	in AFQM. The second is about the noise resilience of AFQM.
	\subsection{Input states.}
		 To ensure a final estimation breaching SQL, (\ref{feedbackAnsatz}) tells us that 
	there should be entanglement between probes in a subsystem  of the total ensemble. Because The ansatz (\ref{feedbackAnsatz}) indicating that the feedback 
	compensations are pushed gradually towards the real parameter value $\phi$.  We need 
	entanglement in a $n$-partite subsystem if we want the feedback compensation $\phi_n$ to be a good estimation of $\phi$ breaching SQL.
	Otherwise, one cannot push $\phi_1,\ldots,\phi_{N-1}$ close enough to $\phi$ to ensure $\phi_{N}$ of violating SQL. The immediate 
	consequence is some highly entangled states such as the GHZ state is not suitable input state in AFQM.
	\subsection{Noise resilience.}
	 Getting close to the final step of AFQM, the feedback 
	compensations would be good estimations of $\phi$ also. If the last few probe qubits are lost, we can still have very a good 
	estimation which is the compensation meant for the lost probes. For example if there are $\ell$ qubits are missing, then 
	the final estimation would be $\phi_{N-\ell}$. Thus our ansatz indicates that there is a built-in resilience of AFQM against 
	\emph{qubit loss}~\cite{Hentschel2010,Hentschel2011,Palittapongarnpim2017b,Palittapongarnpim2019}.
	 In our numerical simulation, we have also considered the \emph{phase fluctuation} noise $\phi_\mathrm{noise}$. 
	} 

	\noindent{In a qubit loss noise channel, a probe qubit has a probability  $\eta$ of being absorbed after entering the channel. Otherwise, the qubit 
	experiences only the driving field $\phi$ without disturbing.
	The phase fluctuation noise can be a random fluctuation added to the parameter field $\phi$ arising from the environment or the imperfection of 
	our feedback control over the qubits. A qubit entering the interferometer would be driven by $\hat{U}_{\phi+\phi_\mathrm{noise}}$ instead of 
	$\hat{U}_\phi$. Typically, we assume such a fluctuation is Gaussian 
	\begin{equation}
		p(\phi_\mathrm{noise}) = \frac{e^{-\phi_\mathrm{noise}^2/\delta\phi_\mathrm{noise}^2}}{\sqrt{2\pi}\delta\phi_\mathrm{noise}}.
	\end{equation}
	Its standard deviation $\delta\phi_\mathrm{noise}$ describes how strong the phase fluctuation is.
	}

%
\section{Combining with Markovian feedback assumption.}  HS indicates that the adjustment of $\phi_n$ from
	the immediate former compensation $\phi_{n-1}$ depends only on the measurement result $s_n$ of $\hat{s}_z^{(n)}$
	\begin{equation}
		\phi_n = \phi_{n-1}-2s_n\Delta_n.
	\end{equation}
	$\Delta_1$,\ldots, and $\Delta_N$ thus constitute the AFQM parameter search
	space~\cite{Hentschel2010,Hentschel2011,Lovett2013,Palittpongarnpim2016,Palittapongarnpim2017,
	Palittapongarnpim2017b,Palittapongarnpim2019,Hayes2014}. By invoking ansatz (\ref{feedbackAnsatz}), we would have
	\begin{equation}
		\phi_n = \phi_{n-1}-2s_n/(n+1)^{\wp_n}
		\,\,\textrm{and} \,\,\,
		\Delta_n =1/(n+1)^{\wp_n}.
	\end{equation}
	The parameter space becomes that of $\wp_1$, \ldots, and $\wp_N$. As has mentioned $|\phi_{n-1}-\phi|$ and $|\phi_{n}-\phi|$ are close, 
	we may expect to see a smooth transition of the slope of $|\phi_n-\phi|$ in a log-log plot of $|\phi_n-\phi|$ versus $n$.
	Many $|\phi_n-\phi|$ would closely follow the same scale $1/n^\alpha$. Thus one  may expect many $\Delta_n$ to closely follow the same scale 
	$1/(n+1)^\wp$. 
	Generally, we expect the adjustment to be a polynomial of the inverse power of $n$
	\begin{equation}
		\Delta_n = \sum_{\ell=0}^{N_\mathrm{s}-1}\frac{c_\ell\pi}{(n+1)^{\wp+\ell}}.
		\label{inverse_scaling_adjustment}
	\end{equation}
	$c_0$, \ldots, and $c_{N_\mathrm{s}-1}$ give us enough flexibility to cope with the deviation of $\Delta_n$ from  the 
	common scale $1/(n+1)^\wp$. $N_\mathrm{s}$ is our choice of number of terms in the expansion.
	(\ref{inverse_scaling_adjustment}) can be seen as a derivative of our feedback ansatz (\ref{feedbackAnsatz}).
	Including $\wp$, there are $N_\mathrm{s}+1$ control parameters,
	the combination of which
	we call an {\it inverse-scaling  policy} $\mathscr{P}$. This reduces the search space dimension to $N_\mathrm{s}+1$
	which is independent of $N$. It enables a reduction of computation complexity. The memory space required to store $\mathscr{P}$ is also
	constant.

%
\section{Devising AFQM via machine learning: cost function and computational complexity.}
Following the HS approach~\cite{Hentschel2010,Hentschel2011,Lovett2013,Palittpongarnpim2016,Palittapongarnpim2017,Palittapongarnpim2017b,Palittapongarnpim2019}, we implement machine learning
	algorithm such as PSO and DE to generate AFQM under the guidance of our ansatz derivative (\ref{inverse_scaling_adjustment}).
	Employing machine learning
	to optimize the AFQM with our feedback ansatz is very much like a treasure hunting under the guidance of SQL and HL.

	\subsection{Cost function.} We employ Holevo variance to quantify the imprecision $\delta\phi$ of
	the final estimation $\phi_N$ as in Ref.~\cite{Holevo2006,{berry2000optimal,berry2001optimal,berry2001adaptive},{Hentschel2010,Hentschel2011,Lovett2013,Palittpongarnpim2016,Palittapongarnpim2017,Palittapongarnpim2017b,Palittapongarnpim2019,Hayes2014}}
	\begin{equation}
		V_\phi = (\delta\phi)^2 = \frac{1}{S^2} - 1,
		\,\textrm{with}\,
		S = \left|\int_{-\pi}^\pi\ud{\phi}P(\phi)e^{i(\phi-\phi_N)}\right|.
	\end{equation}
	Note that $V_\phi$ is an good approximation of the traditional variance in statistics when $\phi_N$ is very close to
	$\phi$~\cite{berry2001adaptive}. One can simulate
	$K=10N^2$ trials of experiment and obtain thus many estimations $\phi_N^{(k)}$. $S$ is the so-called {\it sharpness} and can be estimated via Monte-Carlo method as~\cite{{berry2000optimal,berry2001optimal},{Hentschel2010,Hentschel2011,Lovett2013,Palittpongarnpim2016,Palittapongarnpim2017,Palittapongarnpim2017b,Hayes2014}}
	\begin{equation}
		S = \left|\frac{1}{K}\sum_{k=1}^Ke^{i\left[\phi-\phi_N^{(k)}\right]}\right|.
	\end{equation}

	\subsection{Computational complexity.} Both PSO and DE employ a group
	of searching agents and record the best strategy found by the agents throughout their evolving~\cite{Kennedy1995,Storn1996}.
	With greater number of searching agents, one can find better outcome at the cost of adding computational complexity. By our
	ansatz derivative (\ref{inverse_scaling_adjustment}), we need  $\Xi=20(N_\mathrm{s}+1)$ searching agents instead of
	$20N$~\cite{Hentschel2010,Hentschel2011,Lovett2013,Palittpongarnpim2016,Palittapongarnpim2017,Palittapongarnpim2017b,Palittapongarnpim2019} when the input state  is the 
	so-called sine state {(\ref{sine_state_def})}.
	%
	We have also considered feeding the spin-squeezed state~\cite{kitagawa1993}
	\begin{equation}
		\ket{\psi_\mathrm{sss}} = e^{i\hat{J}_x\delta_\mathrm{adj}}e^{-i\hat{J}_z^2T_\mathrm{s}}\ket{jj}_x
		\,\,\textrm{with}\,\,
		\delta_\mathrm{adj}=\frac{1}{2}\arctan\frac{B}{A},
	\end{equation}
	where $A = 1-\left(\cos2T_\mathrm{s}\right)^{N-2}$ and $B = 4\sin{T_\mathrm{s}}\left(\cos{T_\mathrm{s}}\right)^{N-2}$.
	Adding $T_\mathrm{s}$, the parameter
	space dimension would be $N_\mathrm{s}+2$ and thus we dispatch $\Xi=20(N_\mathrm{s}+2)$ agents to search if SSS has been fed
	to the interferometer.
	Given $N_\mathrm{s}=4$, the search space boundaries has been chosen according to
	Table~\ref{search_space_boundary_table}.
	\begin{table}[ht!]
		\caption{Boundaries of inverse-scaling policy parameters.
		Since we suspect $\wp$ to be very close to the region between $1/2$ (SQL) and $1$ (HL), we choose the search zone that
		covers the region between SQL and HL and $10$ times bigger. The boundaries for $c_0$,\ldots, and $c_{N_\mathrm{s}-1}$ are
	empirical which provides
		good results but not guaranteed to be optimal. We choose the upper bound  $2/\sqrt{N}$ for spin squeezing time $T_\mathrm{s} $
		since $1/\sqrt{N}$ is the
	minimum time needed to ensure  maximal quantum Fisher information of $\ket{\psi_\mathrm{sss}}$~\cite{pezze2009entanglement}.
	Recall that quantum Fisher information quantifies the metrology prowess of $\ket{\psi_\mathrm{sss}}$~\cite{braunstein1994statistical}.
	}
		\begin{ruledtabular}
			\begin{tabular}{ccc}
				$\wp$	& $c_\ell$		 & $T_\mathrm{s}$\\
				$[0,5]$ & $[-5,5]$	& $[0,2/\sqrt{N}]$
			\end{tabular}
		\end{ruledtabular}
		\label{search_space_boundary_table}
	\end{table}
	We iterate both PSO and DE for $N_\mathrm{I}=300$ times as has been done in
	Ref.~\cite{{Hentschel2010,Hentschel2011,Lovett2013,Palittpongarnpim2016,Palittapongarnpim2017,Palittapongarnpim2017b,Palittapongarnpim2019}}. The $N$-partite inverse-scaling policy can be generated directly,
	without knowing any $(N-k)$-partite policy for $1\le{k}\le{N-1}$. To generate a $N$-partite
	inverse-scaling policy we hence need time of $\mathcal{O}(K\Xi{N}^2)=\mathcal{O}(N^4)$.
	Recall that the number $\Xi$ of searching agents
	is constant independent of $N$ while each simulation of the adaptive feedback metrology progress consumes time of
	$N^2$~\cite{Hentschel2011b}.

%
	\section{Results and analysis.} 
\begin{figure}[ht]
		\includegraphics[width=0.5\textwidth]{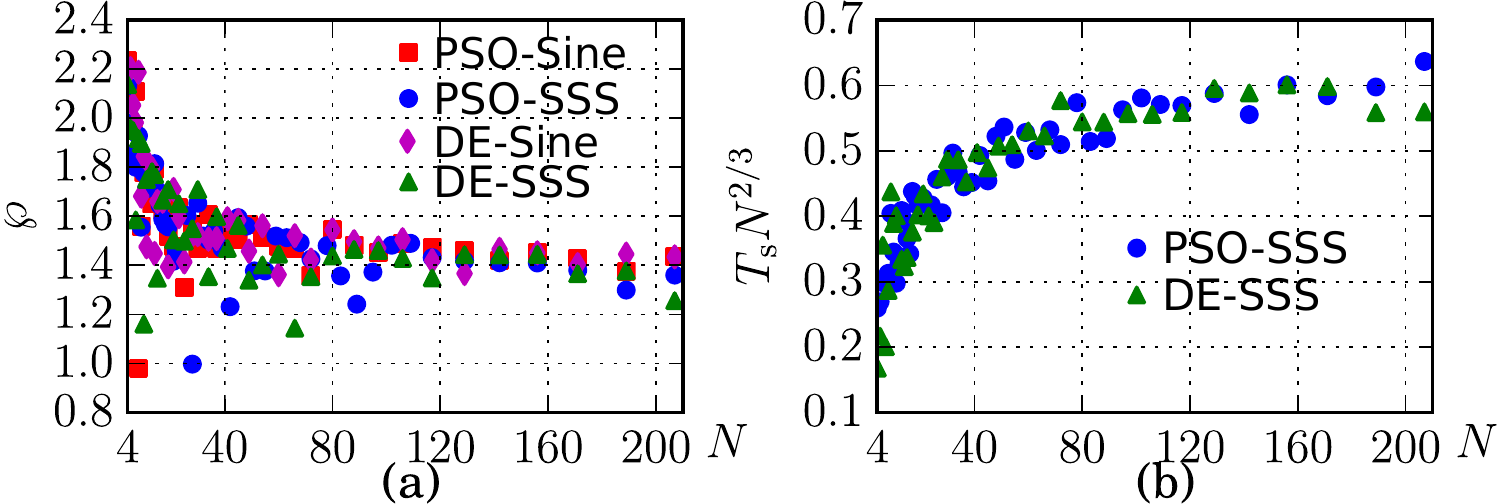}
		\caption{(Color online) General trending of inverse-scaling policy. (a) Leading inverse-scaling exponentiate $\wp$ of the feedback 
		adjustment $\Delta_n$ 
		and (b) optimal squeezing time for AFQM.}
		\label{wp_afqm}
	\end{figure}

We have consider the sine state for comparison with previous results. SSS has been considered due to 
	its well recognized noise-resisting ability and
	proven synthesis procedure in labs~\cite{Hald1999,Fernholz2008,Takano2009,Gross2010,Leroux2010,Hamley2012,Sewell2012,Muessel2014,Hosten2016,Zou2018}. We generate
	AFQM for both kinds of input states via PSO as well as DE. There are thus four groups of data for four different combination of input states 
	and training algorithms which we analyze and present in the following.
\begin{figure*}[ht!]
		\includegraphics[width=\textwidth]{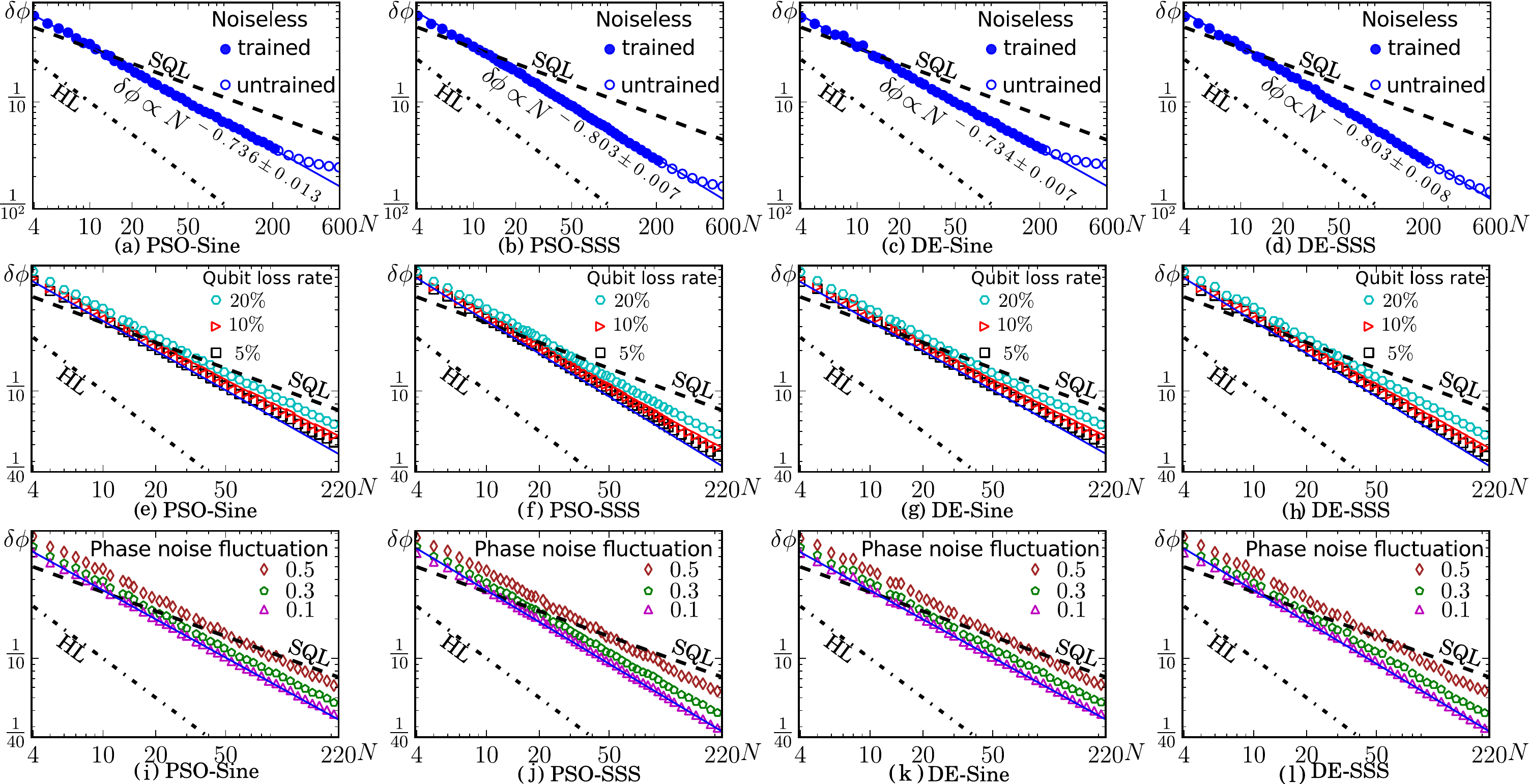}
		\caption{
				(Color online) Imprecision $\delta\phi$ of AFQM generated via machine learning.
				PSO-SSS indicates AFQM with SSS input trained by PSO. Similar nomenclature rule applies to PSO-Sine, DE-SSS and DE-Sine.  
				{All} dots are numerical data while the solid {blue} lines are generated by least-squares fitting the data represented by 
				{blue round-solid} dots.
				{(a-d) Performance in noiseless scenarios.} 
				Every {blue round-solid } dot represents AFQM generated via machine learning while the 
				{blue round-hollow } dots represent AFQM employing
				inverse-scaling policy for $N=207$ without further training.
				{(e-h) Performance of noiseless-channel oriented policy against qubit loss.
				(i-l) Performance  of noiseless-channel oriented policy against phase fluctuation noise.}
			}
		\label{hpv_invSFeedback_sss}
\end{figure*}
	We have summarized {nine} main conclusions drawn from our numerical data. i) AFQM generated with our feedback ansatz is equally good
	as the previous AFQM generated via HS approach. Given sine input state, this is clear from Fig.~\ref{hpv_invSFeedback_sss}(a,c) and
	Table~\ref{search_space_boundary_table}.
	ii) We can obtain inverse-scaling policy for bigger $N$ in shorter time. For example, we have generated the $207$-partite inverse-scaling 
	policy for the sine state via PSO at the cost of approximately $200$ hours running of $120$ CPUs at 2.6 GHz. 	
	iii) Since the parameter space having a much small dimension $N_\mathrm{s}$, PSO and DE produces almost equally good AFQM.
	There is no breakdown of PSO up to $N=207$.
	iv) By optimizing the squeezing time as well, feeding SSS state to the interferometer can
	outperform AFQM with sine input state.
	v) The inverse-scaling policy trained for
	$N=207$ can also sever as a good policy for AFQM with bigger $N$. As shown in Fig.~\ref{hpv_invSFeedback_sss}, the power-law scaling
	of imprecision $\delta\phi$ does not breakdown immediately for $N>207$ if the inverse-scaling policy of $N=207$ is applied without further training. 
	 {Note that all data are plotted 
	in log-log scale in Fig.~\ref{hpv_invSFeedback_sss}. As shown in Fig.~\ref{hpv_invSFeedback_sss}(c), the $207$-partite  policy trained by DE for
	sine state  has moderately 
	good performance for $220$-partite, $260$-partite and $307$-partite systems which correspond to the first three round hollow dots from left to 
	right. In the worst case of our result  if sine state has been employed as shown in Fig.~\ref{hpv_invSFeedback_sss}(c), the $207$-partite policy works very well in the zone 
	of $207{\le}N\lesssim307$ which corresponds to the region between the last blue round-solid dot and the third blue round-hollow dot from left to right in Fig.~\ref{hpv_invSFeedback_sss}(c). For SSS
		which is more stable, the $207$-partite policy works very well in the zone of $207{\le}N\lesssim427$ which correspond to the
		region between the last blue round-solid dot and the fifth blue round-hollow dot from left to right in Fig.~\ref{hpv_invSFeedback_sss}(b).
	} 
	{vi) Our feedback policies have moderate resilience against environment noises such as the qubit loss noise and the phase fluctuation noise. 
	We have tested the policies designed for noiseless scenarios against both the qubit loss and phase fluctuation noise. Given up to $20\%$ of the total
	qubits lost to the environment during the interference, the SQL can still be breached by our policies (cf. Fig.~\ref{hpv_invSFeedback_sss}(c-f)).
	In the case of phase fluctuation up to $0.5$ which is $15.9\%$ of a $\pi$ pulse, violating SQL can still be achieved with our policies
	(cf. Fig.~\ref{hpv_invSFeedback_sss}(e-f)).
	}
	{vii)} As a matter of fact, we can see a general trending of of the leading inverse-scaling exponentiate $\wp$ (cf. Fig.~\ref{wp_afqm}(a)).
	From the fair success of the $207$-partite inverse-scaling policy applying to bigger prob ensembles as well as the general trending
	of $\wp$, one can see the validity and merit of our ansatz (\ref{feedbackAnsatz}) and its derivative (\ref{inverse_scaling_adjustment}).
	vii) In fact we also see a general trending of the optimal spin-squeezing time $T_\mathrm{s}$ (cf. Fig.~\ref{wp_afqm}(b)). For $N$ big enough
	($N{\gtrsim}100$),
	our numerical result suggests that the optimal spin-squeezing time should be approximately $0.6/N^{2/3}$. In fact, we have been optimizing
	$c_\mathrm{s} = T_\mathrm{s}N^{2/3}$ in our simulation.  In applying the $207$-partite inverse-scaling policy for SSS with $N>207$, it is
	$c_\mathrm{s}$ that has been inherited instead of $T_\mathrm{s}$. This hints
	that the optimal squeezing time for employing SSS in AFQM should scale as $1/N^{2/3}$.
	{ix)} As long as the inverse-policy has been trained employing either PSO or DE,  up to $N=207$ the scaling of the imprecision $\delta\phi$ 
	would not break. This upper limit for $N$ would be much bigger, since we can see the scaling persistence in a moderate
	range when the $207$-partite inverse-policy has been applied without training for $N>207$.

	\section{Conclusion and discussion.}
	We have proposed the feedback ansatz (\ref{feedbackAnsatz}) for devising AFQM. When combined
	with the Markovian feedback assumption, we have demonstrated the prowess of our ansatz via numerical simulation.
	{ The policies can be trained in much less time. They have good performance in the noiseless scenario and great resilience 
	against environment noise and experiment imperfections.}
	It may also be useful in devising multi-parameter estimation
	schemes. It is interesting to see that HL and
	SQL can be used as guidelines for AFQM designing instead of being mere metrology performance borderlines. Our method may provide more insight on
	this direction of research.

\begin{acknowledgments}
	We thank B.C. Sanders for stimulating discussions.
	This work was supported National Key R \& D Program of China (Grant Nos. 2016YFA0302104 and 2016YFA0300600), National
	Natural Science Foundation of China (Grant Nos. 11774406 and 11934018), Strategic
Priority Research Program of Chinese Academy of Sciences (Grant No. XDB28000000),
and Beijing Academy of Quantum Information Science (Grant No. Y18G07).
\end{acknowledgments}

\bibliography{Bibliography}
\end{document}